# Nanotribological characterization of industrial Polytetrafluorethylene-based coatings by atomic force microscopy

A. Podestà, G. Fantoni, P. Milani

INFM-Dipartimento di Fisica, Università di Milano,

via Celoria 16, 20133 Milano, Italy.

C. Guida[1], S.Volponi

ABB Service S.r.l. – Research Division

v.le Edison 50, 20099 Sesto San Giovanni, Milano, Italy

**Abstract**

We present the result of a systematic study of the tribological properties of industrial Polytetrafluorethylene (PTFE)-based coatings carried out with an atomic force microscope. A new characterization protocol allowed the reliable and quantitative assessment of the friction coefficient and adhesion forces at the sub-micrometer scale even for highly corrugated industrial samples. We have studied and compared PTFE coatings charged with different additives in dry and humid environment. The influence of additives and humidity on the friction coefficient and on adhesion forces has been investigated using standard silicon nitride tips as sliders in the low-load regime.

---

[1] Present Address: ABB Service S.r.l. - Group Functions, v. Lama 33, 20099 Sesto San Giovanni, Milano, Italy.

**Introduction.**

Friction, wear and adhesion play an important role in determining the performances of components of industrial devices. At present a great interest is devoted to functional and structural coatings providing low friction, high hardness and wear resistance, and other properties that improve the tribological performances of devices [1]. In particular, in the field of dry lubricants, PTFE-based coatings are interesting for their very low friction coefficient and general good resistance to heat and corrosion in different application environments [2].

PTFE based coatings are already extensively used in a variety of industrial applications, such as oil-free bearings and anti-friction, anti-stick components, as for example sensor systems in contact with aggressive chemical fluids [3]. The use of dry lubricants, i.e. oil-free mechanical contacts, would find widespread applications if highly durable and mechanically resistant coatings could be applied even under severe working conditions. The use of fillers and additives in PTFE is the most common way to improve the mechanical properties of the coatings [2].

Macroscopic tribological parameters such as the friction coefficient, material hardness and adhesion strength of the coating are currently characterized with instruments which can investigate length scales ranging from few tens of microns to few tens of centimetres, in different environmental conditions (temperature, humidity, presence of lubricant layers, etc...) [1]. For example, the friction coefficient is typically measured with pin-on-disk and similar techniques, providing a contact area from few tens of squared microns to few tens of squared centimetres, while the hardness can be measured with indenters using diamond tips with standard geometries providing contact areas as small as few hundreds of squared nanometers [1,4].

The macroscopic properties of a material are determined by physico-chemical mechanisms occurring at smaller scales, down to the atomic and molecular ones. The understanding of these phenomena related to the micro and nano-structure of the coatings, and the study of their dependence on both internal (composition, presence of additives, thermal annealing, etc…) and external (relative humidity, ambient temperature, lubricants) parameters would provide a better control of the material properties and performances.

In order to perform quantitative friction measurements down to the nanometer scale it is necessary to control the movement of a nanometer-sized probe in close contact with the sample surface and accurately monitor the forces acting on the probe. At these scales, moreover, the mechanical behaviour is influenced by several parameters that must be taken into account in order to extract quantitative and reproducible results. Among them, the morphological parameters, such as roughness, granularity, power spectrum, play an important role.

The atomic force microscope (AFM) is one of the most suitable instruments for the characterisation of both the morphological and tribological properties of many different types of materials with nanometer resolution in both ambient and controlled environment [5-7]. AFM cantilevers act as very sensitive force sensors, thanks to their force constants as small as 0.005 N/m. Both vertical and lateral forces acting on the tip can be monitored at the same time, thus allowing the acquisition of topographic and lateral force (i.e. friction) maps. This fact makes the AFM a privileged candidate to perform quantitative nano and micro-tribological characterisation of materials having a corrugated surface, whose tribological properties are supposed to be tightly connected to the morphological ones.

In this paper we present the results of a systematic study of the frictional properties of PTFE-based industrial coatings carried out with an atomic force microscope. We have developed a new characterization protocol that provides a reliable value of the parameters affecting the frictional behaviour of corrugated samples in ambient conditions. These parameters are the friction coefficient and the zero-load friction force, which is related to adhesion.

The protocol represents an effort to provide a characterization tool for systems, which are far from ideal conditions such as flat crystalline surfaces in ultra-high vacuum. In particular, the presence of surface roughness and strong adhesion due to capillary forces makes the interpretation of AFM lateral force maps difficult and requires the application of suitable numerical procedure for data analysis.

Friction and other tribological properties are scale dependent. The contact mechanics and consequently the friction regime of the slider-sample interface depend upon the relative size of the two parts [8,9]. The sub-micron tribological properties measured with the AFM are thus not directly comparable to the macroscopic ones. However, as mentioned above, a study at the sub-micron scale may help identifying the effects of physico-chemical and environmental parameters on the tribological performances at larger scales. On the basis of our experimental results we will discuss the influence on the tribological performances of the different compositions used in the coating production process.

**Experimental.**

We have studied four sets of samples: PTFE (polytetrafluoroethylene), PTFE+$MoS_2$, PTFE+polyurethane (PU), and $MoS_2$+Al. They all consist in coated steel plates: the first three are PTFE-based, obtained by thermal treatments of nano-emulsions [10,11], i.e. dispersion of polymer particles of nanoscale size in water; the fourth is a $MoS_2$-based coating obtained by Physical Vapour Deposition (PVD). The first and the fourth samples are reference homogeneous material coatings, while the second and the third are PTFE-based compositions, with $MoS_2$ filler or PU modified PTFE.

We have used a Nanoscope IIIa Multimode AFM from Digital Instruments operated in contact mode. Standard V-shaped silicon nitride cantilevers with integrated square-pyramidal tips were used (tip radius 20-50 nm). We have carefully calibrated the cantilever vertical force constant using the "thermal noise method" described by Butt *et al.* [12]. The measured force constant were typically in the range 0.08-0.12 N/m, with a relative error of about 10%. The cantilever-photodetector z-sensitivity *zsens* was extracted by force vs. distance curves with a relative error of about 10%. The two factors allow to transform the photodetector output voltage $\Delta U_{vert}$ in the applied load $L$: $L = zsens\, k\, \Delta U_{vert}$. We have calibrated the lateral sensitivity $\alpha$, which is the proportionality factor between the lateral force *f* and the lateral voltage signal from the photodetector: $\Delta U_{lat}$: $f=\alpha \Delta U_{lat}$, using the method proposed by Ogletree *et al.* [13], obtaining values of $\alpha$ of the order of 30 nN/V, with a relative error of about 20%. With a dedicated procedure for the topographic correction of AFM lateral force signals, described in detail elsewhere [14], assuming a modified Amonton's law for friction (linear dependence plus offset), we extracted from lateral force maps reliable and statistically strong values of the friction coefficients and adhesion constants. Applied load is remotely controlled through the deflection set-point using a second PC housing a data acquisition board and running software written in LabView environment.

Adhesion force is extracted from force vs. distance curves as the depth of the pull-off region [15]. The range of total applied load (external load plus adhesion) varied from 0 nN to 50 nN, such that we are confident to maintain a wearless regime [16]. Each line of the lateral force map corresponds to a scan at a certain applied load (scan size 1 μm). The load is changed after each line, for a total of 512 lines acquired. Thanks to our protocol we are thus able to record a complete lateral force vs. load ramp in a single 512x512 points AFM image. This allows collecting a good statistics in a few scans. We have carried out friction measurements in wet (RH~45%) and dry (RH<5%) nitrogen at room temperature, using a home-made sealed humidity and atmosphere controlled chamber. The sliding speed in all the friction tests was 2 μm/s.

**Results and discussion**.

*The model and the algorithm.*

We assume the following friction law for the systems under investigation:

$$f=\mu N+C \quad (1)$$

Here *f* is the friction force and *N* is the *total* load acting along the local surface normal, including the extra contribution from attractive adhesive forces (that is, $N=N_{ext}+N_{int}$, where $N_{ext}$ is the external applied load and $N_{int}$ is the surface adhesion). This linear dependence of friction on load is typical of both the plastic and the elastic multi-asperity regime [17-19]. Eq. (1) differs from the well known Amonton's law [19] by the presence of the offset *C*, which represents the residual friction force at zero total applied load. This zero-total load friction force is typical of single-asperity adhesive contacts (like the JKR model [20-22]). Its use in the case of a multi-asperity contact is justified by the fact that in the limit of low loads at the junction between a tiny AFM tip and the sample surface there are only a few asperities in contact, and the regime approaches the single-asperity one. The use of Eq. (1) is discussed in theoretical works [4,23]. The presence of the offset *C* in Eq. (1) is thus related to adhesion in analogy to the adhesive single-asperity case and represents an additional contribution to the term $\mu N_{int}$ in Eq. (1). The stronger is adhesion, the larger *C* is expected to be. Adhesion is an important parameter of friction measurements carried out at the sub-micron scale. Materials with the same friction coefficient should have different adhesion properties and consequently, from Eq. (1), they should have different absolute friction forces under the same applied load.

The typical morphology of a PTFE-based coating is shown in Fig. 1. The topographic map was acquired in contact-mode with the AFM. The surface roughness on a scale of 150 μm is about 500 nm. On the scale of few μm, typical of the friction measurements presented in this paper, the roughness is still of the order of tens of nanometers and the tilted regions extend up to few hundreds of nanometers. The AFM tip radii used in the friction measurements were in general smaller than 50 nm, such that the correction of lateral force maps from the effects of local corrugation is needed [14,24,25]. Notice in Fig. 1 the presence of directional features in the horizontal plane that are the memory of the polishing of the steel substrate.

A typical lateral force vs. external applied load curve measured on a PTFE-based coating obtained from all the points of the lateral force map having a well defined slope in the corresponding topographic map is shown in Fig. 2. The dependence is linear, except for very low loads in the retracting region, where attractive adhesive forces retain the tip while a negative external load is applied. In Fig. 2 the external applied load of $N_{ext}$=–15 nN corresponds to *total* applied load equal to zero. A residual friction force of about 0.3 nN is present. This residual force must not be confused with the zero-*external* applied load (at $N_{ext}$=0 nN) friction force of about 1 nN. The apparent friction coefficient and offset extracted by a linear fit must be topographically corrected to give the true values. This topic is of central importance for FFM experiments on corrugated samples. Actually, in the case of a locally tilted surface, the measured forces in the directions parallel and perpendicular to the AFM reference plane do not necessarily coincide with the forces acting parallel and perpendicularly to the sample surface, which actually define the friction coefficient and the friction vs. load characteristics of the interface under investigation. Assuming a modified Amonton's law for friction (Eq. (1)) we have developed a procedure for carrying out the topographic correction of lateral force maps which is basically a generalization of the one proposed by Bhushan *et al.* [24,25] (our procedure is described in details in Ref. [14]). Considering all the values of the frictional parameters extracted from the curves obtained at different slopes (such as the one shown in Fig. 2) one obtains the histograms of the measured corrected friction coefficients and adhesive constants. An example of such histograms is given in Fig. 3. Finally, the experimental values of both $\mu$ and $C$ and their errors are obtained via a gaussian fit.

*Friction coefficients and adhesion constants.*

In Tabs. 1 and 2 we report the average values of $\mu$ an $C$ of different coatings. As explained in the previous section, each AFM lateral force map provides the values of $\mu$ and $C$ with their errors. The values reported in Tabs. 1 and 2 are weighted averages taken over 5-10 lateral-force maps for each type of coating.

We first consider the values of the friction coefficients measured in humid and dry environment shown in Tab. 1. PTFE and PTFE+MoS$_2$ coatings have similar friction coefficients. The friction coefficients look also independent on relative humidity, at least in the range [0-45]%. A similar trend is observed also for the PTFE+PU sample, while in this case the friction coefficient is larger (the highest value of all coatings). A different behaviour is found for the MoS$_2$+Al sample. In this case the friction coefficient increases in humid environment.

The values of the offset $C$ shown in Tab. 2 are generally small (always less than 1 nN), but still significant with respect to the absolute value of frictional forces (in the range [0-5] nN, see Fig. 2) and definitely different from zero, in the limit of the experimental error. This fact represents an a–posteriori validation of the use of the modified Amonton's law (1) in our analysis. The behaviour of the adhesive constant $C$ is the same for all the samples (see Tab. 2) except for the PTFE+MoS$_2$ coating. In this case the value of the adhesion constant decreases instead of increasing, as one would expect, when humidity is increased.

*Discussion.*

We first notice that the friction coefficient $\mu$ of two samples can be the same but the absolute friction force at a given load varies. This is a consequence of considering explicitly the effect of adhesion, i.e. of including the offset $C$ in the Amonton's law of friction. This difference at the macroscopic scale is likely to be negligible. As a matter of fact adhesion force is usually neglected in macroscopic friction, because it is in general very small compared with the applied load in the Newton range. The adhesion term may play some role however in highly miniaturised mechanical devices, like MEMS and magnetic storage devices [26], where contact areas can be as small as few tens of squared nanometers and the contact and friction regime approaches the one of the Friction Force Microscope. Whatever the effect of adhesion is on the absolute value of friction, the adhesion constant $C$ can give interesting information about the chemistry of the interface under investigation and its dependence on additives and different composition of materials.

Considering first PTFE and PTFE+PU we notice that these samples have a humidity independent friction coefficient and an adhesive offset $C$ that increases (slightly) when humidity increases. The friction coefficient of MoS$_2$+Al increases with humidity. All these materials are worst lubricants in humid environment than in a dry one, the absolute friction force at a given load being larger in humid than in dry environment.

The case of PTFE+$MoS_2$ is somehow anomalous. While the friction coefficient is humidity independent, the adhesive offset $C$ is not: it decreases instead of increase while humidity increases. This is opposite to the behaviour of both $MoS_2$+Al and PTFE. The consequence of this anomalous behaviour is that PTFE+$MoS_2$ has a larger net friction force in dry environment than in the presence of humidity.

On a macroscopic scale $MoS_2$ is a good lubricant in dry ambient and under vacuum. It shows a degradation of performances in the presence of water and oxygen [27,28]. We observe an increase in the friction coefficient of $MoS_2$+Al in humid nitrogen. The addition of $Mos_2$ to PTFE however does not alter the value of the friction coefficient, which is almost insensitive to relative humidity. The adhesive constant $C$ decreases of about 30% increasing the humidity degree, while in pure $MoS_2$ it stays constant and in pure PTFE it increases of about 40%, as one would intuitively expect. Recent AFM studies have shown that in $MoS_2$ films in ambient atmosphere, Van der Waals forces represent the strongest component of the adhesive force (the other being mostly capillary forces, related to the presence of water, and electrostatic forces) [29]. The adhesive properties of $MoS_2$-added materials would thus be expected to be quite insensitive to the water content of the film surface (this does not imply necessarily that the friction coefficient stays constant). The behaviour of $MoS_2$+Al is coherent with this experimental observation: its adhesive constant is independent on humidity. The behaviour of PTFE+$MoS_2$ however is not directly predictable from that of its components. Our results suggest that the PTFE+MoS$_2$ coating behave as a different material with respect to both PTFE and MoS$_2$. In particular, adhesion is inhibited, thanks to the non-hydrophilic properties of MoS$_2$, while the friction coefficient is not affected with respect to pure PTFE. PTFE+MoS$_2$ is thus a good lubricant, especially in humid environment. These results show that the addition of MoS$_2$ to PTFE improve the structural resistance of the film [30] and also the tribological performances by lowering the absolute friction force and the adhesion with other surfaces.

**Conclusions.**

We have carried out a quantitative characterization of frictional properties of industrial PTFE-coatings via atomic force microscopy, on a sub-micron scale. Assuming a linear dependence of friction on normal load following a modified Amonton's law (friction coefficient plus adhesive offset) we have measured the friction coefficients and adhesive constants of the coatings in humid and dry environment. We have found a weak dependence of friction coefficients on the relative humidity, except for the MoS$_2$+Al coatings, which has a larger friction coefficient in the presence of humidity.

We have observed that the dependence of the adhesive constant on humidity in the case of PTFE+MoS$_2$ is anomalous: it decreases by increasing humidity. This causes the PTFE+MoS$_2$ coating to have better structural stability and tribological properties than pure PTFE, especially in dry environment.

References.

Table captions.

Table 1. Measured friction coefficients $\mu$ of PTFE-based coatings in humid and dry Nitrogen.

Table 2. Measured adhesive constants $C$ of PTFE-based coatings in humid and dry Nitrogen.

Figure captions.

Figure 1. Typical AFM topographic map of a PTFE-based coating. The AFM was operated in contact-mode. Scan size is 130 µm and vertical scale is 3 µm.

Figure 2. Typical lateral force vs. external applied load curve for a PTFE-based coating. The dotted line is a weighted linear fit of experimental data.

Figure 3. Typical histograms of friction coefficients and adhesion constants measured on PTFE-based coatings. From such histograms average values and standard deviations are extracted via a gaussian fit (not shown).

Table 1.

| FRICTION COEFFICIENT $\mu$ | | |
|---|---|---|
| **SAMPLE** | **DRY NITROGEN** | **WET** |
| PTFE | $0.032 \pm 3\cdot10^{-3}$ | $0.034 \pm 3\cdot10^{-3}$ |
| PTFE+$MoS_2$ | $0.030 \pm 3\cdot10^{-3}$ | $0.031 \pm 3\cdot10^{-3}$ |
| $MoS_2$+Al | $0.039 \pm 4\cdot10^{-3}$ | $0.053 \pm 5\cdot10^{-3}$ |
| PTFE+PU | $0.064 \pm 6\cdot10^{-3}$ | $0.067 \pm 7\cdot10^{-3}$ |

Table 2.

| OFFSET $C$ [nN] | | |
|---|---|---|
| **SAMPLE** | **DRY NITROGEN** | **WET** |
| PTFE | $0.27 \pm 3 \cdot 10^{-2}$ | $0.38 \pm 4 \cdot 10^{-2}$ |
| PTFE+MoS$_2$ | $0.22 \pm 2 \cdot 10^{-2}$ | $0.15 \pm 1 \cdot 10^{-2}$ |
| MoS$_2$+Al | $0.36 \pm 4 \cdot 10^{-2}$ | $0.39 \pm 4 \cdot 10^{-2}$ |
| PTFE+PU | $0.083 \pm 8 \cdot 10^{-3}$ | $0.13 \pm 1 \cdot 10^{-2}$ |

Figure 1.

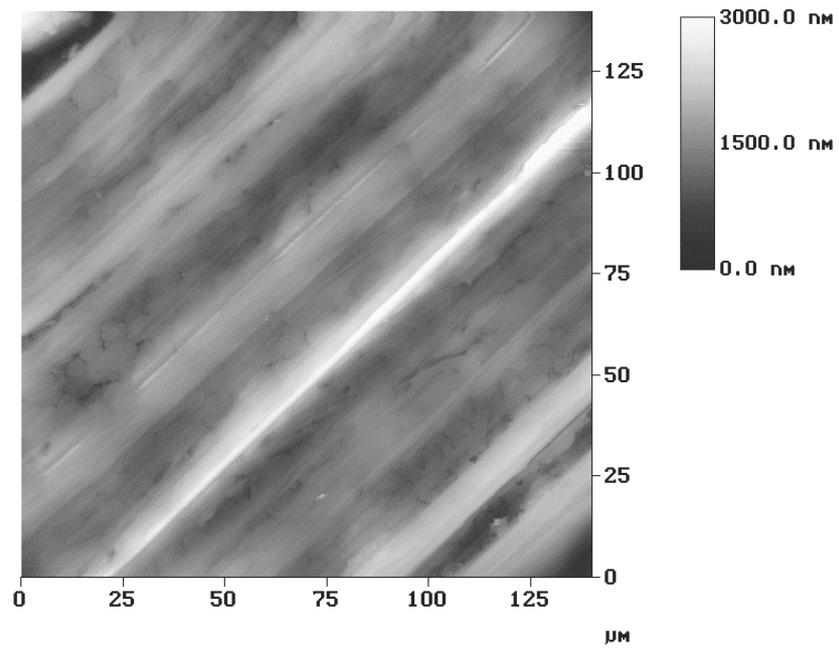

Figure 2.

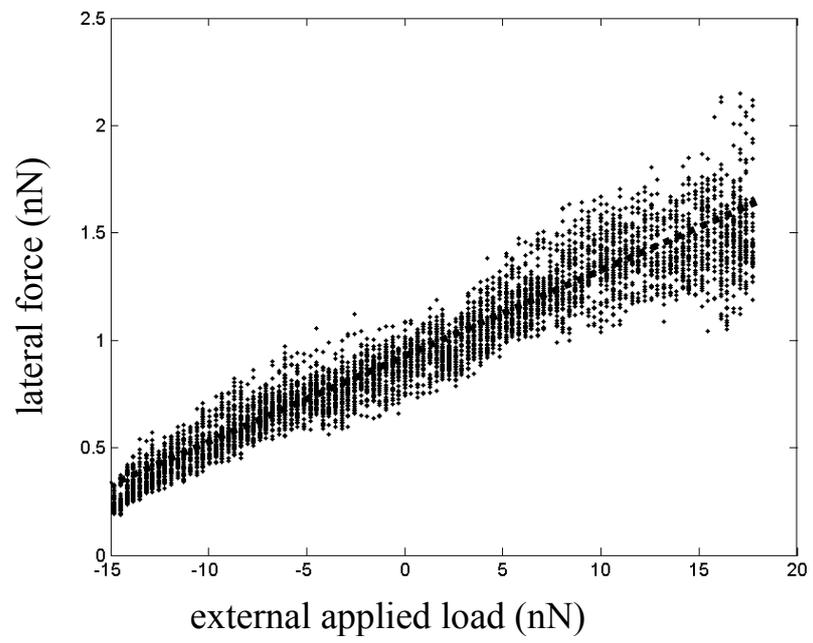

Figure 3.

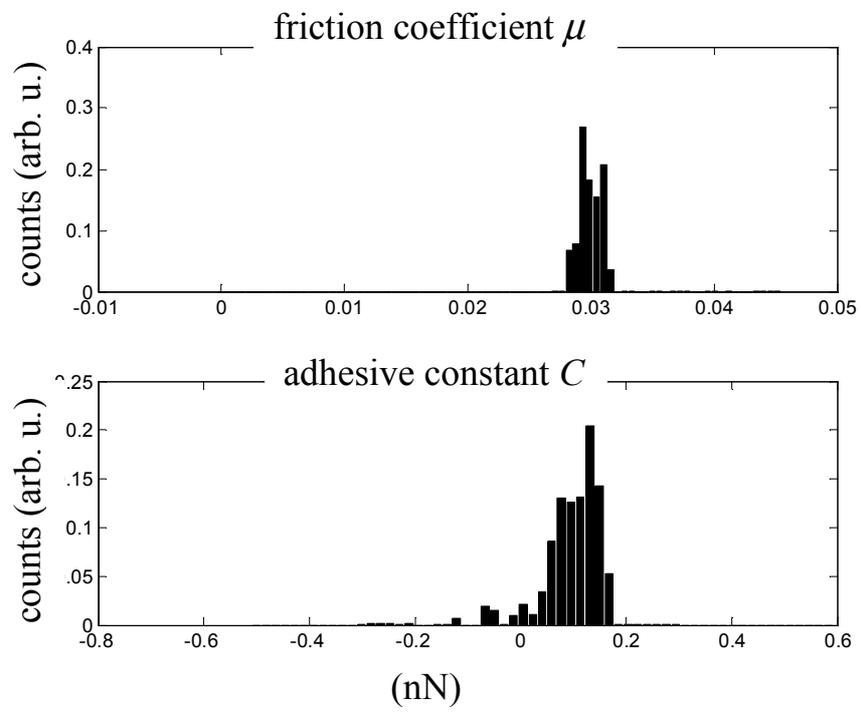